\documentclass [10pt,twocolumn,a4paper] {article}
\usepackage{epsfig,color}

\newcommand{\beq}{\begin{equation}}
\newcommand{\eeq[1]}{\label{#1} \end{equation}}
\newcommand{\bear}{\begin{eqnarray}}
\newcommand{\eear[1]}{\label{#1}\end{eqnarray}}
\newcommand{\foot}{\footnote }

\begin{document}

\title{{\Large \bf Effects of nucleus initialization on event-by-event observables.}}
\author{{\large \ B.~Mattos~Tavares$^{a}$ \foot{\em  bernardo@if.ufrj.br}, 
H.-J. Drescher$^{b}$, T. Kodama$^{a}$}
\\  \\{\em \small  $^{a}$Instituto de F\'{\i}sica, Universidade Federal do Rio de 
Janeiro, 68528, 
21945-970}\\ {\em \small Rio de Janeiro, RJ, Brasil}
\\  \\{\em \small  $^{b}$ Frankfurt Institute for Advanced Studies (FIAS), 
Johann Wolfgang Goethe-Universi\"at,}\\
{\em \small  Max-von-Laue-Str. 1, 60438 Frankfurt am Main, Germany}}

\date{\small (September 15$^{th}$,  2006)}

\maketitle
%
%%%%%%%%%%%%%%%%%%%%%%%%%%%%%%%%%%%%%%%%%%%%%%%%%%%%
% ===================       WITH CORRECTIONS FROM HAJO !!       =========================
%%%%%%%%%%%%%%%%%%%%%%%%%%%%%%%%%%%%%%%%%%%%%%%%%%%%
%
\abstract{\small{In this work we present a study of the influence of
nucleus initializations on the event-by-event elliptic flow
coefficient, $v_2$. In most Monte-Carlo models, the initial positions of
the nucleons in a nucleus are completely uncorrelated, which can lead to
very high density regions. In a simple, yet more realistic model 
where overlapping of the nucleons is avoided, fluctuations in the
initial conditions are reduced. However, $v_2$
distributions are not very sensitive to the initialization choice.}}
%
%%%%%%%%%%%%%%%%%%%%%%%%%%%%%%%%%%%%%%%%%%%%%%%%%%%%
\section{Introduction}

The main goal of studying ultra-relativistic heavy ion reactions is to
identify (or not) the formation of the Quark-Gluon Plasma
(QGP). Event-by-event (EbyE) analysis of global observables can
significantly contribute in this direction \cite{voloshinfluc}. One
way to measure EbyE fluctuations is the statistical approach: each
observable in an event should contain fluctuations and the
distribution of such an observable can be characterized by its mean
value and higher moments \cite{kochfluc}.

The most interesting fluctuations are dynamical, since they give
important information about the formation of the system.  Possible
examples are the occurrence of jets, giving rise to fluctuations in
the high $p_t$ tail of transverse momentum distributions, and the
fluctuation in the anisotropic flow coefficient $v_2$ due to unusual
hard/soft equation of state or fluctuating initial conditions
\cite{voloshinfluc, tkodamaBJP35}.

Other sources of fluctuations of statistical or technical nature can
also be present in EbyE distributions. For example,
finite multiplicity affects the determination of $\langle p_t
\rangle$, ratios of multiplicities of particle species and also the
strength of the anisotropic flow coefficient, due to imprecise
determination of event plane \cite{voloshinfluc}.
In this work, we are interested in a technical source of
fluctuations, commonly present in event generators, the initialization
of the positions of the nucleons in a nucleus before the collision.
%
%%%%%%%%%%%%%%%%%%%%%%%%%%%%%%%%%%%%%%%%%%%%%%%%%%%%%%%%%%%%%
\section{Hydrodynamical model}

We have proposed that dynamical and technical sources of fluctuations
can be studied in the framework of relativistic hydrodynamics, which
allows one to separate the relevant physics from statistical noise
\cite{teseber}.  If the local thermal equilibrium is attained in
relativistic heavy ion collisions, hydrodynamical description may be
most adequate for the space-time evolution of the system. Once the
initial conditions (spatial configuration of 4-velocity field and
conserved currents) of the system are specified, 
%
%the subsequent
%dynamics is determined by the equation of state of the matter.  
%
the principal factor which characterizes the hydrodynamical motion is the
equation of state of the matter. 
When the
hydrodynamical prescription is not valid anymore, one must employ a
decoupling criterion to generate particles, i.e, hadrons. So we have
only three inputs in the hydrodynamical approach: initial conditions,
equation of state and the decoupling criterion.

In this work, we employ the numerical code SPHERIO \cite{spherio},
which solves the hydrodynamical equations of motion in 3 dimensions
and can deal with any kind of spatial configuration in the initial
conditions. For the equation of state, we have adopted a quark-gluon
free gas with a bag constant of 380 MeV/fm$^3$ for the QGP phase, and
a hadron resonance gas with excluded volume (only for baryons) in the
confined phase.  Almost all the resonances up to 2.5 GeV have been used.  
The phase boundary is obtained via the
Gibbs criterion, and the transition is first order for every chemical
potential.  As a decoupling criterion, we adopted Cooper-Frye
procedure \cite{cooperfrye} with a fixed freeze-out temperature of
140~MeV. For further details in the equation of state and decoupling
prescription used, see \cite{osadaranp}.

%
%%%%%%%%%%%%%%%%%%%%%%%%%%%%%%%%%%%%%%%%%%%%%%%%%%%%%%%%%%%%%
\section{Nucleus initialization: correlated versus uncorrelated 
nucleon positions.}

For the initial condition, we use the NEXUS event generator
\cite{nexus}. Based on the Gribov-Regge model of hadronic collisions,
it generates a spatial distribution of the energy-momentum tensor
$T^{\mu\nu} $ and the baryon number density $n_B$ on the hyper-surface
$\tau = const$.  Monte-Carlo generation of events from this model give
rise to physical fluctuations in the initial conditions.  However, the
procedure chosen to initialize the nuclei before the collision, may
generate also unphysical fluctuations, which can influence EbyE
observables.
%
%%%%%%%%%%%%%%%%%%%%%%%%%%%%%%%%%
%\subsection{Correlated versus uncorrelated nucleon positions}

Usually one employs the Wood-Saxon distribution
\beq
\rho(r) = \rho_0\frac{1}{1 + exp[(r - R)/D]},
\eeq[woodsax_results]
to determine the initial position of the $A$ nucleons within a
nucleus. The constants in eq. (\ref{woodsax_results}) are the nuclear
density $\rho_0 = 0.16$~fm$^{-3}$, the nucleus radius $R$ and the
diffuseness parameter D $\sim 0.55$ fm (usually).

The nucleons determined this way are completely uncorrelated, like a
gas of free particles. So, it can happen that several nucleons occupy
the same position in space, which is unphysical. However, this method
is used in many Monte-Carlo codes, since it is widely believed that
unphysical fluctuations average out, at least for inclusive spectra.
Some models, e.g HIJING \cite{hijing} are aware of this problem, and
employ a rejection technique for nucleons which are closer to each
other than a given distance. However, for realistic distances (in the
order of 2 times the proton radius $\sim$ 1.6 fm) this leads to a
pushing of nucleon positions towards the shell and lowers density in
the center. Nexus generates nuclei, with completely uncorelated
positions.

\begin{figure}[!h]
\begin{center}
\epsfig{file=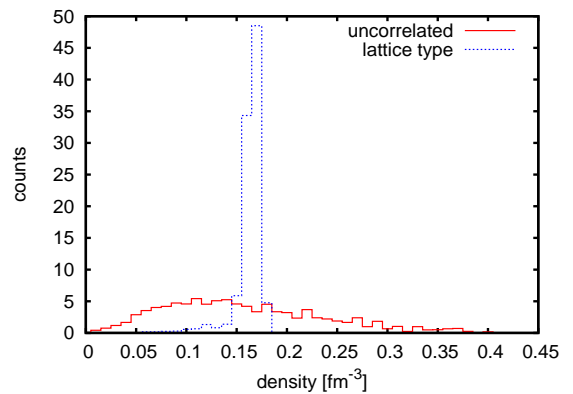,width=7.5cm,angle=0} 
\caption{Distribution of the nucleon density in the center of the nucleus ($r<4$~fm), 
for correlated (dashed line) and uncorrelated (solid line) nucleons.}
\label{nucldens_fig}
\end{center}
\end{figure}

The most suitable initialization of a nucleus would be a selection of
nucleon positions according to a realistic wave function which
includes correlations \cite{lfrankfurt}. Each nucleon would be
surrounded by a corelation hole \cite{lfrankfurt}, which prevents
other nucleons from occupying the same space.  For simplicity, we
employ a lattice model in order to demonstrate the effect. The nucleons 
are placed on a body-centered cubic lattice (BCC) and the
positions are then accepted with a probability $\rho/\rho_0$ given by
eq.  (\ref{woodsax_results}). The packing efficiency of a BCC lattice
is $e=8/3*\pi/(4/\sqrt{3})^3=0.68$ and the lattice spacing is
$l=(e/\rho_0/\pi*3/4)^{1/3}*4/\sqrt{3}=2.32~$fm \cite{kittel}. This method correctly
reproduces the density in the center of the nucleus, as can be seen in
Fig. \ref{nucldens_fig}. To avoid lattice artifacts, each nucleus is
rotated by some random angles.  We shall call this method lattice (or
correlated) initialization.
%
%\foot{A better choice would be an FCC lattice, that guarantee a spheric distribution.}
%
%%%%%%   DISTRIBUICAO DE DENSIDADE DE NUCLEONS   %%%%%%

\begin{figure}[!h]
\begin{center}
\epsfig{file=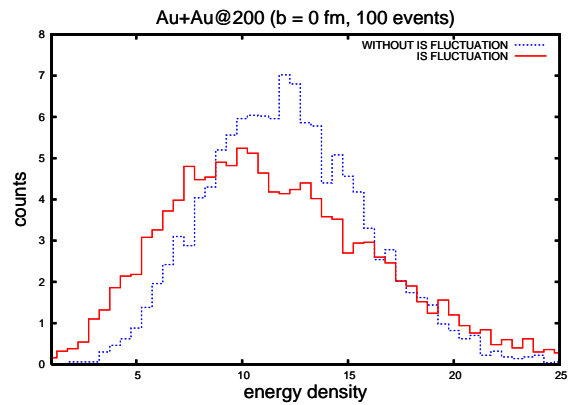,width=7.5cm,angle=0}
\caption{Energy density distribution for central collisions (b = 0).}
\label{edens_fig}
\end{center}
\end{figure}

In Fig. \ref{edens_fig} we show a plot for the energy density
distribution for the correlated (dashed line) and the uncorrelated (full
line) cases. Each distribution represents 100 NEXUS events for
Gold-Gold collisions with $\sqrt{s} = 200$ GeV per nucleon pair, with
zero impact parameter $b$ (central collisions).  One can easily check
that the energy density from correlated nucleons is narrower than the
uncorrelated case, as expected. It can also be seen that the average
energy density is a  larger for the lattice type initialization. 
%
%%%%%%%%%%%  DISTRIBUICAO DE ENERGIA  %%%%%%%%%%%
%
%%%%%%%%%%%%%%%%%%%%%%%%%%%%%%%%%%%%%%%%%%%%%%%%%%%%%%%%%%%%%
\section{Results}

\begin{figure}[!h]
\begin{center}
\epsfig{file=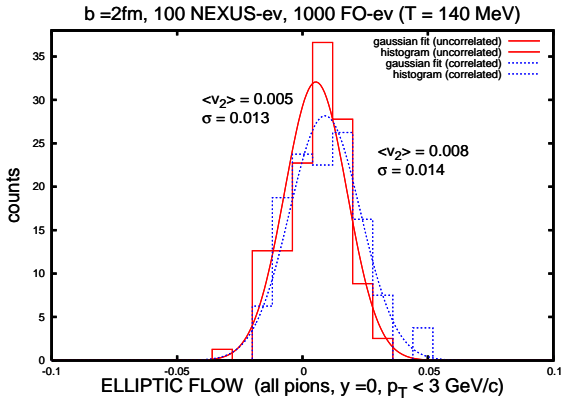, width=7.5cm,angle=0}
\caption{Elliptic flow distributions for central collisions ($b$ = 2 fm). }
\label{v2b2_fig}
\end{center}
\end{figure}

\begin{figure}[!h]
\begin{center}
\epsfig{file=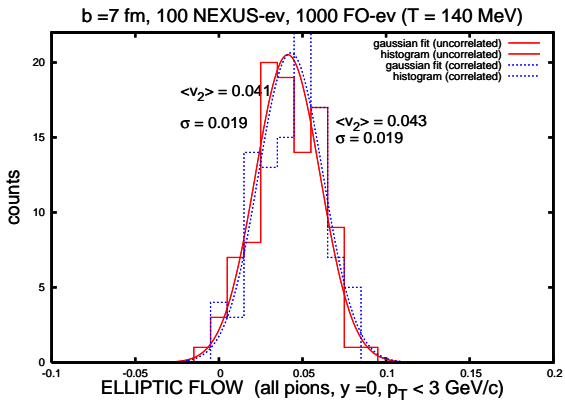, width=7.5cm,angle=0}
\caption{Elliptic flow distributions for mid-central collisions ($b$ = 7 fm). }
\label{v2b7_fig}
\end{center}
\end{figure}

\begin{figure}[!h]
\begin{center}
\epsfig{file=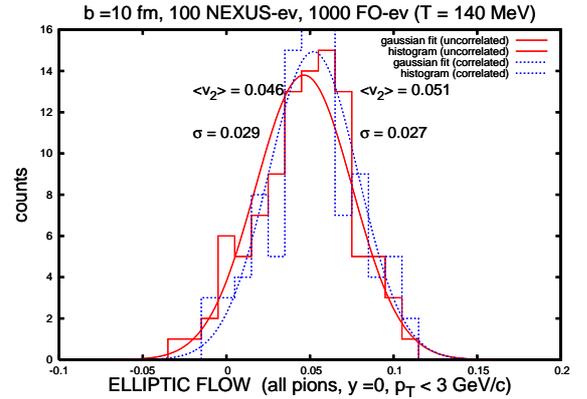,width=7.5cm,angle=0}
\caption{Elliptic flow distributions for peripheral collisions ($b$ = 10 fm). }
\label{v2b10_fig}
\end{center}
\end{figure}

In Figs. \ref{v2b2_fig}, \ref{v2b7_fig} and \ref{v2b10_fig} we show
EbyE distributions for the elliptic flow coefficient $v_2$
(integrated over $p_t$) of pions, computed for $b$ = 2~fm (central
collisions), $b$ = 7~fm (mid-central) and $b$ = 10 fm (peripheral
collisions), at mid-rapidity, for Gold-Gold reactions at $\sqrt{s} =
200$ GeV per nucleon pair. We show the histograms and
the Gaussian fits for the observable originated from the correlated
(dashed) and uncorrelated (solid) nucleus initialization. We computed
100 events per type of initialization at each impact parameter.
Despite the differences in the energy density distributions, the $v_2$
in all centralities shows very similar variances $\sigma$. That may be
caused by the hydrodynamical expansion which reduces the effects of the
fluctuations. 
%
%%%%%%%%%%%  DISTRIBUICAO DE V2  %%%%%%%%%%%%%%
%%%%%%%%%%%%%%%%%%%%%%%%%%%%%%%%%%%%%
%
%%%%%%%%%%%%%%%%%%%%%%%%%%%%%%%%%%%%%%%%%%%%%%%%%%%%%%%%%%%%%
\section{Conclusions and perspectives}

In this work we studied the effects of nucleus initialization on EbyE
 observables. The $v_2$ parameter is not very sensitive to the
 choice of nucleus initialization. The influence on other observables 
are in progress.  We also plan to study the effects of
 impact parameter distribution, reaction plane determination (for the
 $v_2$ case) and other finite multiplicity related observables. The
 goal is to develop a method that separates physical from statistical
 fluctuations.

\paragraph{Acknowledgments:} 
the authors would like to thanks for discussions with A. Dumitru. This
work has been done with the support of FAPESP, CNPq, DAAD and CAPES.

%
%%%%%%%%%%%%%%%%%%%%%%%%%%%%%%%%%%%%%%%%%%%%%%%%%%%%%%%%%%%%%

%
%%%%%%%%%%%%%%%%%%%%%%%%%%%%%%%%%%%%%%%%%%%%%%%%%%%%%%%%%%%%%
%\section{Figures}

%
%%%%%%%%%%%%%%%%%%%%%%%%%%%%%%%%%%%%%%%%%%%%%%%%%%%%%%%%
\end{document}